\documentclass{article}
\usepackage{epsfig}
\newcommand{\bfr}{\begin{flushright}}
\newcommand{\efr}{\end{flushright}}

\begin{document}
\title{The LHCf experiment: \\
present status and physics results
\thanks{Presented at  the Low x workshop, June 13-18 2017, Bari, Italy}%
}

\author{E. Berti\footnote{eugenio.berti@fi.infn.it}
\\
INFN Section and University of Florence, Italy\\
{\small
 on behalf of the \textit{LHCf} collaboration
}
\smallskip\\
}


\date{\today}

\maketitle
\begin{abstract} 
The main aim of the LHCf experiment is to provide precise measurements of the production spectra relative to neutral particle produced by high energy proton-ion collisions in the very forward region. This information is necessary in order to test and tune hadronic interaction models used by ground-based cosmic rays experiments. In order to reach this goal, LHCf makes use of two small sampling calorimeters installed in the LHC tunnel at $\pm 140$ m from IP1, able to detect neutral particles having pseudo-rapidity $\eta > 8.4$.  In this paper we will present the current status of the LHCf experiment, regarding in particular the first analysis results from data taking relative to p-p collisions at $\sqrt{s} = $ 13 TeV.
\\
~
\\
PACS number(s): 
13.85.-t, 13.85.Tp
\end{abstract}

\section{Introduction}
In order to understand the processes responsible for acceleration and propagation of cosmic rays in the universe, measurements of their flux and composition up to the Greisen-Zatsepin-Kuzmin cut-off (GZK cut-off) are necessary. These measurements are performed by ground-based experiments through the indirect detection of the extensive air showers (EASs) that cosmic rays form when interacting with the atmosphere. The properties of the primary particle are then reconstructed making use of MC simulations that, being EASs physics described by soft (non perturbative) QCD, necessarily rely on phenomenological models. Among them very different predictions are found at high energies, due to the lack of experimental calibration data, resulting in large systematic uncertainties on cosmic rays measurements. The main purpose of the LHC forward (LHCf) experiment is to provide important information for the calibration of hadronic interaction models in the forward region. The Large Hadron Collider (LHC) is the most suitable place where to perform these measurements, because a center of mass energy of $\sqrt{s} = $ 13 TeV in p-p collisions is equivalent to about $9 \times 10^{16}$ eV in the reference system where the target is at rest, an energy not so distant from the one of Ultra High Energy Cosmic Rays (UHECRs).

\section{The experiment}
LHCf \cite{ref:lhcf} consists of two small sampling calorimeters installed in the Large Hadron Collider (LHC  \cite{ref:lhc}) tunnel at $\pm 140$ m from IP1 (ATLAS \cite{ref:atlas} interaction point). Being placed after the D1 dipole magnet, only neutral particles produced by proton-ion collisions and having pseudo-rapidity $\eta>8.4$ can reach the experiment. Each one of the two detectors, called Arm1 and Arm2, is made up by two square towers of 22 W and 16 GSO (plastic scintillator before 2014 upgrade) layers for a total length of 29 cm, equivalent to 44 $X_{0}$ and 1.6 $\lambda_{I}$. Towers size is 20 mm $\times$ 20 mm and 40 mm $\times$ 40 mm for Arm1, 25 mm $\times$ 25 mm and 32 mm $\times$ 32 mm for Arm2. Energy resolution is better than $5\%$ for $\gamma$s above 100 GeV and about $40\%$ for hadrons above 500 GeV. The transverse position of the incident particle is reconstructed using 4 xy imaging layers inserted at different depths. They are formed by 1 mm width GSO-bars (scintillating fibers before 2014 upgrade) in the case of Arm1 and by 160 $\mu$m read-out pitch silicon microstrip detectors in the case of Arm2. Position resolution is better than 200 $\mu$m for $\gamma$s above 100 GeV and 1 mm for hadrons above 500 GeV. More detailed descriptions of the detector are reported elsewhere \cite{ref:lhcf_photon_performances, ref:lhcf_neutron_performances}.

\section{Analysis results}

Because LHCf requires low luminosity and high $\beta ^{*}$, so far data have been acquired during special runs: in 2009-2010 p-p collisions at $\sqrt{s} = $ 0.9 and 7 TeV, in 2013 p-p at $\sqrt{s} = $ 2.56 TeV and p-Pb at $\sqrt{s_{NN}} = $ 5.02 TeV, in 2015 p-p at $\sqrt{s} = $ 13 TeV, in 2016 p-Pb at $\sqrt{s_{NN}} = $ 5.02 and 8.1 TeV. In this paper we will discuss about the main analysis results obtained in LHC Run I and, in particular, about the ongoing activity relative to data acquired in LHC Run II.
\\
\begin{figure}[!t]
\centering
\includegraphics[width=\linewidth]{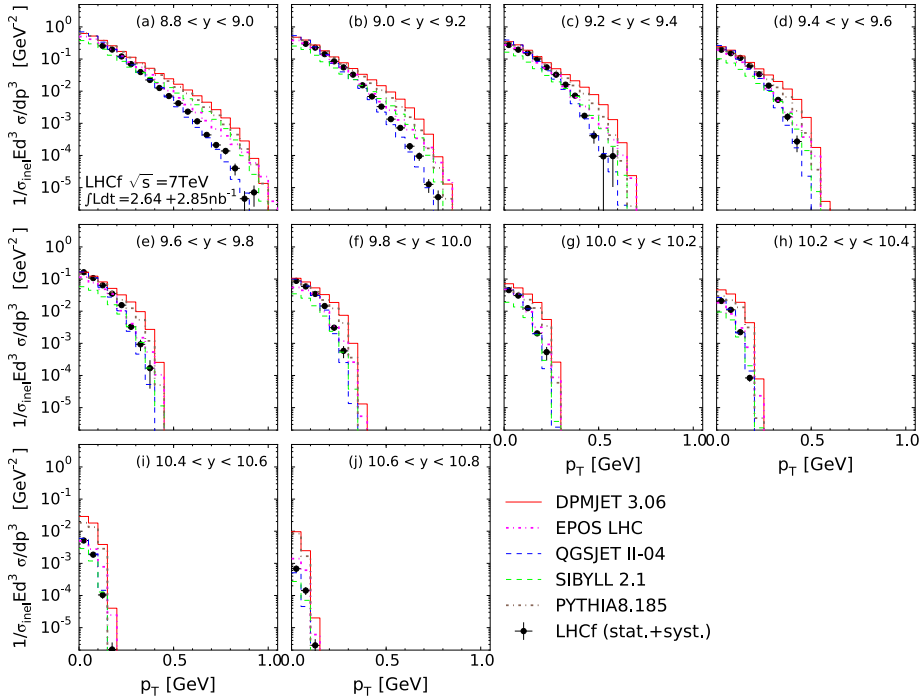}
\caption{Inclusive production cross section of neutral pion produced in p-p collisions at $\sqrt{s} = $ 7 TeV expressed as a function of $\mathrm{p_{T}}$ in different $y$ regions \cite{ref:piall}. Filled circles with error bars represent experimental results with the total statistical and systematic uncertainty, whereas lines with different colors refers to the predictions relative to several hadronic interaction models.}
\label{pt7tev}
\end{figure}
\newline
One of the most significant results achieved from Run I data is the measurement of inclusive production cross section of forward neutral pions \cite{ref:pi7tev, ref:pi5.02tev, ref:piall}, indirectly reconstructed from the detection of the two $\gamma$s originated in the decay. There are at least two reasons for this importance. The first reason is that $\pi^0$ playes an essential role in EASs evolution, where it transfers energy from the hadronic to the electromagnetic channel: in this sense, LHCf measurements of $\mathrm{p_{T}}$ ($\mathrm{p_{Z}}$) spectra as a function of $y$ ($\mathrm{p_{T}}$) have shown that QGSJet II-04 \cite{ref:qgsjet} is the model with the best overall agreement, whereas the other generators does not satisfactorily reproduce the  experimental results in all the regions considered in the analysis (see for example Fig.\ref{pt7tev}). The second reason is that, being the detector optimized for the reconstruction of electromagnetic showers, $\pi^0$ is the most powerful probe we have to investigate about general properties of EASs physics: in this sense, thanks to data relative to p-p and p-Pb collisions, LHCf have measured the nuclear modification factor, enlightening a suppression of production spectra by a factor of 0.1--0.2 in the case of lead target respect to proton target, and, thanks to data relative to different collisions energy, LHCf have tested several scaling laws ($\mathrm{<p_{T}>}$ scaling \cite{ref:pt}, Feynman scaling \cite{ref:feynman}, limiting fragmentation \cite{ref:fragmentation1, ref:fragmentation2, ref:fragmentation3}), all of them generally holding at a 10-20\% level. 
\\
\newline
\enlargethispage{+1\baselineskip}
The activity on data taken in Run II has started extending the analyses runned on p-p collisions at $\sqrt{s} = $ 7 TeV to the 13 TeV case. In particular we focused on the energy spectra of forward photons and neutrons produced in the collisions. In both cases we considered different $\eta$ regions, defined in such a way to be common on both Arm1 and Arm2 detectors. After estimating all correction factors and all systematic uncertainties, we performed bayesian unfolding \cite{ref:dagostini} in order to minimize the effect of detector response in our final measurement.

\begin{figure}
\centering
\includegraphics[width=0.9\linewidth]{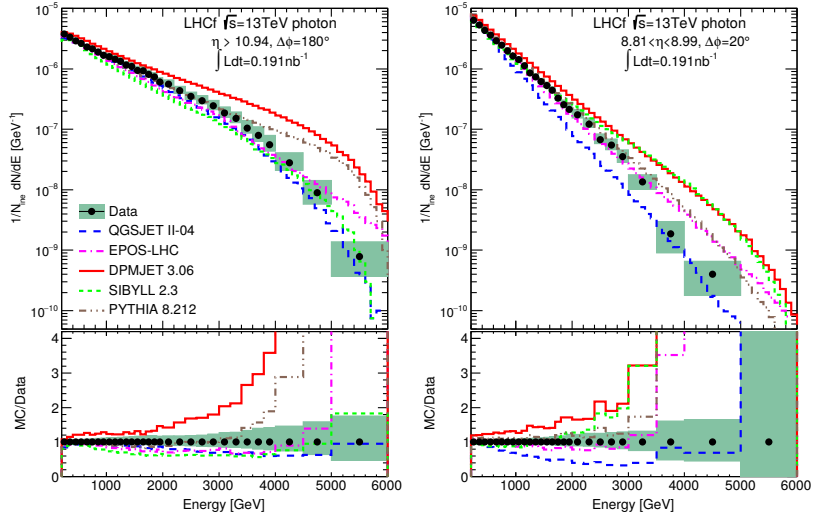}
\caption{Photon energy spectra obtained from the combination of LHCf Arm1 and Arm2 results compared with MC predictions \cite{ref:photon13tev}. The top panels show the energy spectra, and the bottom panels show the ratio of MC predictions to the data. The hatched areas indicate the total uncertainties of experimental data including the statistical and the systematic uncertainties.}
\label{photon_13TeV}
\end{figure}

The photons analysis has been completed and the paper is waiting for the final editor review. The inclusive  photon energy spectra are shown in Fig.\ref{photon_13TeV}, where we can see, in a similar way to what was observed in the case of 7 TeV \cite{ref:photon7tev}, that a large variation among different models is present and no one is able to reproduce satisfactorily experimental data in all the energy range  \cite{ref:photon13tev}: QGSJet II-04 agrees rather well with data in $\eta>10.94$, but is softer $8.81<\eta<8.99$; EPOS-LHC \cite{ref:epos} agrees rather well below about 5 and 3 TeV respectively, but is harder at high energy. Even if they both are post-LHC models, there is no strong change in the mechanisms responsible for forward photon production respect to the pre-LHC version used in \cite{ref:photon7tev}. Therefore the observed differences between the 7 and 13 TeV case may correspond to the different $\mathrm{p_{T}}$ coverage.

\begin{figure}[t]
\centering
\includegraphics[width=1.\linewidth]{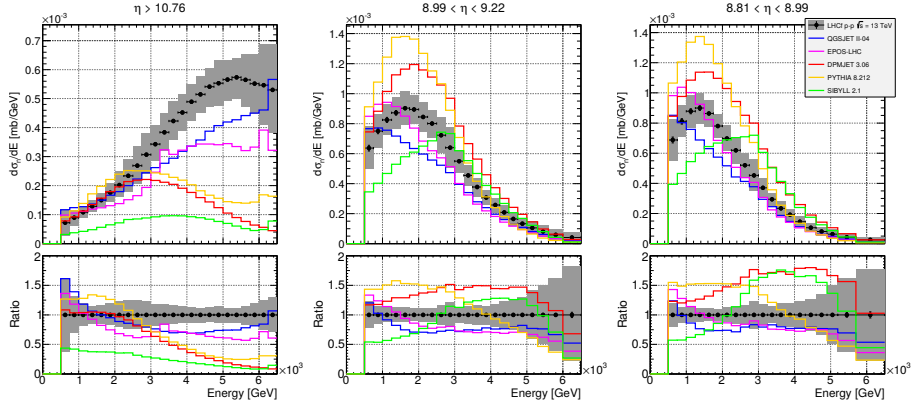}
\caption{Preliminary result relative to the differential neutron production cross section obtained from Arm2 detector compared with MC predictions \cite{ref:neutron13tev}. The top panels show the energy spectra, and the bottom panels show the ratio of MC predictions to the data.  The hatched areas indicate the total uncertainties of experimental data including the statistical and the systematic uncertainties.}
\label{neutron_13TeV}
\end{figure}

The hadron analysis is still ongoing. It is almost completed only for Arm2 for which we already have a preliminary result, shown in Fig.\ref{neutron_13TeV}, relative to the differential neutron production cross section \cite{ref:neutron13tev}. As already observed in the 7 TeV case \cite{ref:neutron7tev}, a very large discrepancy between experimental measurements and model predictions is present in $\eta>10.76$, qualitatively explained only by QGSJet II-04. This fact may have important consequences in cosmic ray physics, because the strong underestimation of neutron production rate at high energy indicates that, in this pseudorapidity region, all generators overestimate the inelasticity, an important parameter in EASs evolution. In $8.81<\eta<9.22$ the agreement of models is generally better, especially in the case of EPOS-LHC.

\section{Conclusions and future prospects}
The LHCf experiment showed that in the forward region no model perfectly reproduces the experimental observations. Measurements of energy, $\mathrm{p_{T}}$ and $\mathrm{p_{Z}}$ spectra of the neutral particles produced in the very forward region can therefore be used to tune these models. The collaboration is now extending the analyses relative to data acquired in p-p collisions at $\sqrt{s} = $ 7 TeV to the 13 TeV case. In parallel, the ATLAS-LHCf common analysis, based on the common operation the two experiments had in 2015 and 2016, is in progress. This is probably the most promising development in the impact of LHCf measurements because, among several benefits that both experiments can have, the ATLAS information can separate diffractive from non-diffractive events observed in Arm1 and Arm2. In addition, two other important analyses are starting: the one on data relative to p-Pb collisions at $\sqrt{s_{NN}} = $ 8.1 TeV acquired in 2016 at LHC and the one on data relative to p-p collisions at $\sqrt{s} = $ 510 GeV acquired in 2017 at the Relativistic Heavy Ion Collider (RHIC \cite{ref:rhic}). Regarding future runs, the LHCf experiment remains very interested in high energy p-O and O-O collisions if at some time they will be available at LHC.

\newpage


\bibliographystyle{apsrev4-1}


\end{document}